\documentclass[twocolumn,showpacs,preprintnumbers,amsmath,amssymb]{revtex4}

\usepackage{epsf}
\usepackage{rotating}

\usepackage{graphicx,epsfig}
\usepackage{dcolumn}
\usepackage{bm}

\newcommand{\req}[1]{Eq.~(\ref{#1})}
\newcommand{\avg}[1]{\langle #1\rangle}

\newcommand{\fig}[1]{Fig.~\ref{#1}}
\newcommand{\tab}[1]{Table \ref{#1}}
\def\qm{Q_{\rm max}}

\def\re{{\rm o}}
\def\rh{{\rho}}

\def\piatp1{{\pi_{\alpha}(t+1)}}
\def\Kat0a{{K_{\alpha}(t_\alpha^0)}}
\def\fa{{f_\alpha}}
\def\fb{{f_\beta}}
\def\qci{{Q_c^{({\rm I})}}}
\def\qcii{{Q_c^{({\rm II})}}}
\def\tre{{\omega}}
\def\sen{{s}}
\def\gen{{g}}
\def\newm{{\gamma}}

\begin{document}

\preprint{}

\title[Title]
{Dynamics underlying Box-office: Movie Competition on Recommender Systems}

\author{C. H. Yeung$^{1,2}$, G. Cimini$^1$ and C.-H. Jin$^1$}
\affiliation{$^1$Department of Physics, University of Fribourg, CH-1700 Fribourg, Switzerland\\
$^2$Research Center for Complex System Science, University of Shanghai for Science and Technology, 200093 Shanghai, P.R. China}

\date{\today}

\begin{abstract}
We introduce a simple model to study movie competition in the recommender systems.
Movies of heterogeneous quality compete against each other through viewers' reviews
and generate interesting dynamics of box-office.
By assuming mean-field interactions between the competing movies,
we show that run-away effect of popularity spreading is triggered by defeating the average
review score,
leading to hits in box-office.
The average review score thus characterizes the critical movie quality necessary for transition
from box-office bombs to blockbusters.
The major factors affecting the critical review score are examined.
By iterating the mean-field dynamical equations,
we obtain qualitative agreements with simulations and real systems in the
dynamical forms of box-office,
revealing the significant role of competition in understanding box-office dynamics.
\end{abstract}

\maketitle


\section{Introduction}

Dynamics underlying movie box-office has been an important issue for both academic and commercial interest
in the past decades of years \cite{devany1996, devany2001, chiou2008}.
As compared to the traditional advertising campaign,
recommender systems nowadays provide reviews and scores which constitute
a platform for movies to compete among 
each other  directly.
Due to the popularity of the Internet,
such competitions play an increasingly significant role in driving movie box-office.
Despite their importance,
only single movie dynamics is considered in conventional approaches \cite{edwards2001, weisbuch2000},
leaving competitions through recommender reviews unattained.
Such movie interactions are in particular interesting 
to understand the physics of movie competitions in influencing box-office dynamics.

Box-office dynamics has been studied using different approaches ranging from
statistics to neural networks.
In early approaches,
potential viewers consider the past total success of a movie \cite{devany1996},
or the decision of its predecessor \cite{devany2001},
to decide whether to watch the movie. 
Though multiple movies are considered in these approaches,
competition though modern recommender systems is not addressed.
Movie competitions are considered in \cite{chiou2008}
but dynamics of the competing movies is not examined.
On the other hand,
single movie dynamics is studied by differential equations \cite{edwards2001}
and the spread of movie awareness 
by automata or percolation \cite{weisbuch2000}.
To understand and predict the box-office dynamics, 
empirical studies are conducted \cite{duan2008, duan2008b}
and statistics based forecast \cite{litman1998} and neural networks  \cite{sharda2006, lee2009} are employed.
Some of these approaches consider
individual factors such as nations,
genre and star values,
and may overlook the importance of competitions among movies.

In this paper,
we introduce a model in which movies compete with each other through 
reviews posted on a recommender system.
Viewers post their reviews 
and drive other potential viewers to watch the movie,
which in turn produce new reviews driving another group of viewers.
The present reviews driven mechanism spread movie popularity and 
generate interesting dynamics in movie box-office.
Here we consider movie reviews as both {\it indicators} as well as {\it influencers}
of box-office,
as suggested by empirical data in \cite{duan2008}.
To capture only the essential elements,
movies in our model are only differentiated by their quality 
and time of introduction.
As different from approaches
which consider heterogeneous viewers,
we assume that 
potential movie goers are homogeneous
and are only driven to watch movies either by movie reviews or movie freshness.
All these ingredients constitute a simple model which facilitate 
the illustration of physical phenomenon behind movie competitions.

We will show that,
by mean-field approximation,
the average review score characterizes the critical movie quality necessary for booms in movie box-office,
and corresponds to a transition from box-office bombs to blockbusters.
The physical reason behind the booms is the success of the movies in 
spreading popularity through the recommender systems,
creating cascades and dynamical hits after their introduction.
Though we are not able to provide an accurate estimate of the average review score,
we show that the analytical results
have quantitative agreements with simulations and real data,
suggesting the present model in describing the fundamentals of 
movie competitions.
Finally,
we generalize the mean-field approximation to analyze the competitions
of two good movies and show that
box-office dynamics of the competing movies are anti-correlated.

The paper is organized as follows.
We describe the formulation of the model in Section \ref{sec_model}.
In Section \ref{sec_mean},
we employ the mean-field approximation for movie interaction
and discuss the dependence of gross box-office and box-office dynamics on the quality of movies.
The competition between two good movies is discussed in Section \ref{sec_comp}.
We finally compare our approximation with simulation and empirical results in Section \ref{sec_sim}.
Conclusions are given in Section \ref{sec_con}.

\section{Model Formulation}
\label{sec_model}

We consider a community of $N$ agents
aiming to watch movies of high quality.
At each time step,
a fraction $p$ of the agents are chosen to be the
{\it potential viewers},
out of which 
a fraction $\tre$, 
which we call the {\it trendiness},
intend to watch a {\it new} movie.
A new movie is introduced at each step.
Each movie $\alpha$,
introduced at time  $t_\alpha$,
is characterized by quality $Q_\alpha$ randomly drawn from the distribution $\rho(Q_\alpha)$.
Starting from the second step after their introduction,
movies are considered to be {\it old} and are recommended
to agents by a centralized recommender system,
which is in the form of a {\it review list} as shown in \tab{tab_review}.
The remaining potential viewers (who have no intention to watch the new movie)
select an old movie from the list.
Thus, 
a high $\tre$ also corresponds to a small dependence 
of movie goers on the reviews.
For a particular movie $\alpha$,
we denote the number of viewers,
i.e. the box-office revenues,
at time $t$ to be $k_\alpha(t)$.

In reality,
potential viewers of a new movie sense the quality of movie from advertisements.
We thus assume that they watch the new movie $\alpha$ with probability $\pi_\alpha\propto Q_\alpha^\sen$,
such that the ``opening" $k_\alpha(t_\alpha)\propto Np\tre Q_\alpha^\sen$.
A suitable proportionality constant would be $\qm^{-\sen}$,
where $Q\le \qm$ as restricted by $\rho(Q)$.
When $\sen=0$,
potential viewers watch the new movie regardless of its quality.
We set $\sen>0$ when agents have a good sense of movie qualities
to make potential viewers inclined to movies of high quality.
Thus,
we call $\sen$ the {\it quality sensitivity}.

On the other hand,
potential viewers of old movies decide to watch an old movie
by gathering
information from websites of movie reviews,
movie magazines
or word-of-mouth recommendations from peer viewers.
We express this kind of centralized recommendations by a
list of movie reviews as shown in \tab{tab_review}.
Movie popularity thus spread among the agents through the list.
For simplicity,
the reviews are expressed in the form of scalar scores.
At a particular time,
we denote the total number of reviews on the list as $L$
and reviews are labeled by $r=1,\dots,L$.
The movie and its corresponding score on the $l$-th reviews
are denoted respectively as $m_r$ and $u_r$.
For a potential viewer $i$,
the probability to choose a movie $\alpha$ from the list is
\begin{eqnarray}
\label{eq_viewP}
	\pi_{i,\alpha}=\frac{\sum_{r=1}^L (1-a_{i, m_r})u_r\delta_{m_r, \alpha}}{\sum_{r=1}^L (1-a_{i, m_r})u_r},
\end{eqnarray}
where $a_{i,\alpha}=1$ if viewer $i$ has already watched movie $\alpha$ and otherwise 0.
\req{eq_viewP} characterizes the competition of on-list movies.
Only reviews from the previous step are shown on the list,
i.e. reviews before the previous step are {\it deleted}.
The up-to-date reviews lead to the natural evolution of box-office.
It can be shown that by this clearing mechanism,
the total number of viewer for a movie is independent of its time of introduction 
(given that the observed time is longer than the lifespan of the movie),
as different from the first-mover effect in citation of scientific papers \cite{newman2009}.
This is of particular importance for the recommendation of objects
where in the long run freshness is important.

As viewers may provide generous or harsh critics,
we assume that they post their reviews with probability $\eta_\alpha\propto Q_\alpha^\gen$.
When $\gen=0$,
they review the movie regardless of its quality.
The volume of reviews is thus an indicator of the box-office.
When $\gen>0$,
they tend to give good comments and review movie of high quality.
By contrast,
$\gen<0$ represents an opposite phenomenon as
reviewers tend to give bad comments,
making bad movies more popular (by number of reviews)
than good movies in the recommender system.
We thus interpret $\gen$ to be the {\it generosity} of the viewers.
In this paper,
we define {\it popularity} of a movie to be the fraction of reviews on the movie (over all current reviews).

To further simplify the model,
we combine \req{eq_viewP} and the generosity,
and assume that the score $u_\alpha$ of a movie is simply its quality $Q_\alpha$,
which implies that all agents in the model are rational and homogeneous.
Note that in this case $u_\alpha$ is continuous and can be considered
as the average estimated quality by the reviewers. 
\req{eq_viewP} thus becomes
\begin{eqnarray}
\label{eq_viewPsim}
	\pi_{i,\alpha}=\frac{\sum_{r=1}^L (1-a_{i, m_r})Q_{m_r}^{\gen+1}\delta_{m_r, \alpha}}{\sum_{r=1}^L (1-a_{i, m_r})Q_{m_r}^{\gen+1}}.
\end{eqnarray}
When $\gen=-1$,
viewers select a movie merely by popularity on the list,
regardless of quality,
which is suggested by the results of empirical study in \cite{duan2008}.

\begin{table}\centering
\vspace{0.5cm}
\begin{tabular}{cc}
\hline
Movie ID & Movie Score \\
\hline 
$\alpha$ & 5 \\
$\newm$ & 3 \\
$\beta$ & 4 \\
$\gamma$ & 2 \\
$\alpha$ & 5\\
$\vdots$ & $\vdots$\\
\hline
\end{tabular}
\caption{Examples of movie reviews
on the recommender system.
}
\label{tab_review}
\end{table}

\section{Box Office Dynamics - The Mean-field Approximation}
\label{sec_mean}

We start to investigate the box-office dynamics of movies by mean-field approximation.
As mentioned,
movies compete with each other by interaction through
the review list,
as potential viewers select one of the on-list movies
by comparing their quality and popularity.
In the mean-field approximation,
we assume a mean interaction between movies
and denote $\avg{u}_\re$ to be the average score over
reviews of on-list movies,
which have been introduced for at least two steps
on the recommender system.
In other words,
at time $t$ with $m_r$ denoting the $r$-th movie on the list,
$\avg{u}_\re$ is given by
\begin{eqnarray}
	\avg{u}_\re=\frac{\sum_{r=1}^{L}\theta(t-t_{m_r}+2)u_{m_r}}{\sum_{r=1}^{L}\theta(t-t_{m_r}+2)}
\end{eqnarray}
where the heaviside function $\theta(y)=1$ for $y\ge 0$ and otherwise 0.
As we have set the review score to be the movie quality,
$\avg{u}_\re =\avg{Q}_\re$.
We then approximate
\req{eq_viewPsim} for movie $\alpha$ introduced at time $t_\alpha$ by
\begin{widetext}
\begin{eqnarray}
\label{eq_meanpi}
	\pi_{i,\alpha}(t)=
	\begin{cases}
	\displaystyle(1-a_{i,\alpha})\frac{k_\alpha(t_\alpha)Q_\alpha^{\gen+1}}{k_\alpha(t_\alpha)Q_\alpha^{\gen+1}
	+Np(1-\tre)\avg{Q^{\gen+1}}_\re}
	&\mbox{for $t=t_\alpha+1$,}
	\\
	\displaystyle(1-a_{i,\alpha})\frac{k_\alpha(t-1)Q_\alpha^{\gen+1}}{k_\alpha(t-1)Q_\alpha^{\gen+1}
	+k_\newm(t_\newm)Q_\newm^{\gen+1}+[Np(1-\tre)-k_\alpha(t-1)]\avg{Q^{\gen+1}}_\re} 
	&\mbox{for $t>t_\alpha+1$,}
	\end{cases}
\end{eqnarray}
\end{widetext}
where movie $\newm$ corresponds to the movie being introduced at time $t_\newm=t-1$.
Readers may argue that 
the average of $\pi_{i, \alpha}(t)$ instead of $Q_\alpha$ would be a better approximation than \req{eq_meanpi}.
Nevertheless,
we keep the present form of approximation as it greatly simplifies the analysis,
while capturing the essential features 
and facilitating physical interpretations.
Justification is given in Section \ref{sec_sim} and the discrepancy between the present approach and the 
simulation results is described.

From \req{eq_meanpi},
we write down the iterative equations which describe the box-office dynamics $k_\alpha(t)$.
To simplify the notation,
we define $f_\alpha(t) = k_\alpha(t+t_\alpha)/Np$,
corresponding to its {\it popularity}.
Thus,
popularity of an on-list movie is proportional to its box-office at every step.
The initial popularity $f_\alpha(0)$ is proportional to $\tre Q_\alpha^\sen$.
We further approximate $\avg{Q^{\gen+1}}_\re$ by $\avg{Q}^{\gen+1}_\re$,
which turns out to be a good approximation 
as the review list
is usually dominated by several good movies
which have similar quality.
The expression for subsequent $f_\alpha(t)$ is given by
\begin{widetext}
\begin{eqnarray}
\label{eq_fa}
	f_\alpha(t) = 
	\begin{cases}
	\displaystyle(1-\tre)[1-p\fa(0)]\left[1+\frac{1-\tre}{\fa(0)}\left(\frac{\avg{Q}_\re}{Q_\alpha}\right)^{\gen+1}\right]^{-1}
	&\mbox{for $t=t_\alpha+1$,}
	\\
	\displaystyle(1-\tre)\left[1-p\sum_{t'=0}^{t-1}\fa(t')\right]
	\left[1+\frac{f_\newm(0)}{\fa(t-1)}\left(\frac{Q_\newm}{Q_\alpha}\right)^{\gen+1}
	+\left(\frac{1-\tre}{\fa(t-1)}-1\right)\left(\frac{\avg{Q}_\re}{Q_\alpha}\right)^{\gen+1}\right]^{-1}
	&\mbox{for $t>t_\alpha+1$.}
	\end{cases}
\end{eqnarray}
\end{widetext}
These equations can be iterated numerically to generate the dynamics
of box-office.

\subsection{Booms in gross box-office}

To obtain the relation between gross box-office and movie quality,
we adopt again the mean-field approximation for interactions with new movies,
and approximate $f_\newm(0)Q_\newm^{\gen+1}$ by $\tre\avg{Q^{1+\sen+\gen}}_\rh/\qm^\sen$, 
which is the value averaged over the distribution $\rho(Q)$ in
the relation $f_\newm(0)\propto \tre Q_\newm^\sen$,
subject to the proportionality constant $\qm^{-\sen}$.
The quantity $\avg{Q}_\re$ is dependent on the dynamics of box-office
and competition between on-list movies,
and hence is difficult to compute.
Nevertheless,
we approximate $\avg{Q}_\re$ by 
\begin{eqnarray}
\label{eq_avgQapprox}
	\avg{Q}_\re\approx\frac{\int dQ \rho(Q) Q^{\sen+\gen+2}}{\int dQ \rho(Q) Q^{\sen+\gen+1}},
\end{eqnarray}
where $Q^{\sen+\gen}$ is proportional to the probability for the movie
to appear on the recommender list after its introduction.
We do not claim that the above expression is a good approximation of $\avg{Q}_\re$,
but we will see that a rough estimate of $\avg{Q}_\re$ is sufficient to generate the features of the model.
We then iterate \req{eq_fa} numerically to obtain $\sum_{t}f_\alpha(t)$,
which is proportional to the gross box-office $K_\alpha$.

\begin{figure}
\centerline{\epsfig{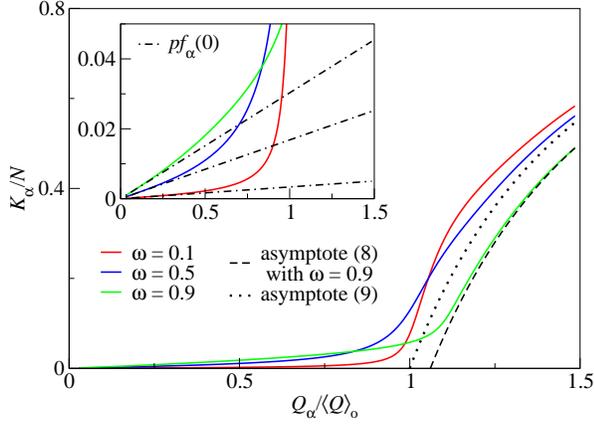}}
\caption{
(Color online) The gross box-office of a movie $\alpha$ 
as a function of $Q_\alpha$. 
Parameters: $p=0.05$, $\sen=1$ and $\gen=1$.
$\avg{Q}_\re$ are obtained with $\rho(Q)\sim Q^{-2}$ with $\qm=50$.
Inset: The same graph in magnified vertical scale. 
}
\label{fig_kq}
\end{figure}

The rescaled gross box-office $K_\alpha/N$
is shown in \fig{fig_kq} as a function of $Q_\alpha$.
The gross box-office of a movie shows a drastic rise,
i.e. a {\it boom},
when its quality is beyond the average review score $\avg{Q}_\re$ of existing competitors.
$\avg{Q}_\re$ thus characterizes the critical quality of movies to become blockbusters.
We note that different forms of $\rho(Q)$ alter $\avg{Q}_\re$,
but not the general picture of booms.
Remarkably,
the gross box-office does not show a large dependence on $\tre$,
the ratio of agents who intend to watch a new movie.
However, 
we will see that the dynamics of $\fa$ does show a dependence on $\tre$.

To get a better understanding of \fig{fig_kq},
we obtain an explicit asymptotic form of the gross box-office
by considering \req{eq_fa} in the large $t$ limit.
In this case,
$O(\fa(t))\ll 1$ and the terms with $\fa^{-1}(t-1)$ becomes dominant in the denominator.
We thus ignores terms of $O(1)$ and rewrite \req{eq_fa} as
\begin{eqnarray}
	\fa(t) = \frac{ (1-\tre)\left[1-p\sum_{t'=0}^{t-1}\fa(t')\right]\fa(t-1)Q_\alpha^{\gen+1} }{ f_\newm(0)Q_\newm^{\gen+1} + (1-\tre)\avg{Q}^{\gen+1}_\re }.
\end{eqnarray}
For movies of low quality,
$\sum_{t}f_\alpha(t)$ is negligible and $\fa(t)$ follows an exponential decay.
For movies of high quality,
their popularity on the recommender system is long lasting,
and $\fa(t)\approx\fa(t-1)$ when $t$ is large.
The gross box-office of the movie is thus given by
\begin{eqnarray}
\label{eq_grossApprox}
	\frac{K_\alpha}{N}
	&\approx& 1-\frac{ \tre\avg{Q_\newm^{\sen+\gen+1}}_\rh/\qm^s + (1-\tre)\avg{Q}^{\gen+1}_\re }
	{(1-\tre)Q_\alpha^{\gen+1}}
	\\
\label{eq_grossApprox2}
	&\approx& 1-\left(\frac{ \avg{Q}_\re }
	{Q_\alpha}\right)^{\gen+1},
\end{eqnarray} 
where the last line is valid only when the second term dominates in the numerator,
i.e. the average on-list movie quality is high such that 
$\avg{Q}^{\gen+1}_\re\gg\avg{Q^{\sen+\gen+1}}_\rh/\qm^s$.
We thus see that when on-list movies are of high quality,
the gross box-office is only weakly dependent on $\tre$ through $\avg{Q}_\re$.
As $Q_\alpha\rightarrow\infty$, 
$K_\alpha/N\rightarrow 1$.
From \fig{fig_kq} and \req{eq_grossApprox},
we can estimate the critical quality $Q_c$ for box-office boom
to be
\begin{eqnarray}
\label{eq_Qc}
	Q_c\approx\avg{Q}_\re,
\end{eqnarray}
which implies blockbusters appear at lower quality
given that competitors are bad movies.

To obtain the degree distribution of movies,
we examine $\sum_{t}f_\alpha(t)$ for movies with low $Q_\alpha$.
As shown in the inset of \fig{fig_kq},
$K_\alpha/N\approx p\fa(0)$ for movies with very low quality,
due to the fast decay of their popularity after introduction.
Assuming movie quality is distributed as $\rho(Q)\sim Q^{-\gamma}$,
the distribution of gross box-office $K$ is given by 
\begin{eqnarray}
\label{eq_pK}
	P(K) \sim K^{-\frac{\gamma+\sen-1}{\sen}},
\end{eqnarray}
which is valid for movies of low quality.
$P(K)$ shows a long tail for large $K$,
due to the boom in gross box-office
for $Q>Q_c$.

\subsection{Hits in box-office dynamics}
\label{sec_dyn}

\begin{figure}
\centerline{\epsfig{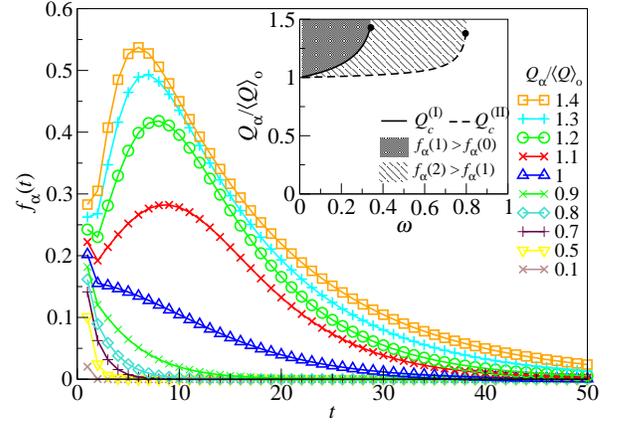}}
\caption{
(Color online) $\fa(t)$ as a function of $t$ for movies of 
various quality $Q_\alpha$.
Parameters: $p=0.05$, $\tre=0.3$, $\sen=1$ and $\gen=1$.
$\avg{Q}_\re$ are obtained with $\rho(Q)\sim Q^{-2}$ with $\qm=50$.
Inset: The phase diagram for hits in box-office.
The critical quality $\qci$ and $\qcii$
as shown respectively by the solid and dashed lines,
as a function of $\tre$.
}
\label{fig_ft}
\end{figure}

To examine the box-office dynamics,
we show in \fig{fig_ft} how $\fa(t)$ evolves with time.
For movies of high quality,
their popularity increases after their introduction,
corresponding to an immediate {\it hit} in box-office 
(see for instance the line with $Q_\alpha/\avg{Q}_\re=1.4$).
We thus consider $\fa(1)>\fa(0)$ which implies
$Q_\alpha$ greater than a critical quality $\qci$,
given by the positive root of equation
\begin{eqnarray}
\label{eq_qc1}
	&&\tre(1-p\tre+p)\left(\frac{\qci}{\avg{Q}_\re}\right)^{\sen+\gen+1} 
	\nonumber\\
	&&-(1-\tre)\left(\frac{\qm}{\avg{Q}_\re}\right)^\sen\left(\frac{\qci}{\avg{Q}_\re}\right)^{\gen+1}
	+(1-\tre)\left(\frac{\qm}{\avg{Q}_\re}\right)^\sen
	\nonumber\\
	&&= 0.
\end{eqnarray}
The corresponding $\qci/\avg{Q}_\re$ for different values of $\tre$ are 
shown by the solid line in the inset of  \fig{fig_ft},
which implies that immediate hit occurs for movies with quality in the shaded region.
Though explicit solution for $\qci$ is difficult to obtain,
we see from \req{eq_qc1} that when $\sen=0$,
only the rescaled quantity $\qci/\avg{Q}_\re$ is relevant.
Thus,
defeating existing movies on the reviews list
is a major physical origin for an immediate hit in the box-office dynamics.
Moreover,
a hit is more likely to occur
at time with bad movies in the list,
i.e with low $\avg{Q}_\re$.

The solution of $\qci$ exists only when the trendiness is smaller than some threshold $\tre_c$
given by the root of the following equation
\begin{eqnarray}
\label{eq_trec}
	-\Xi p\tre_c^2+[\Xi(1+p)+1]\tre_c-1 = 0,
\end{eqnarray}
when $\sen=\gen=1$,
where $\Xi=3\sqrt{3}\avg{Q}_\re^\sen/2Q_{\rm max}$.
Beyond the threshold,
an immediate hit does not occur regardless of movie quality.
It implies that this phenomenon is less likely to occur 
when viewers have high intention to watch new movies.
From \req{eq_trec},
we see that $\tre_c$ increases with decreasing $\avg{Q}_\re$,
implying that an immediate hits occur at high trendiness
when on-list movies are bad.

Another interesting behavior is observed
in the range of $\qcii<Q_\alpha<\qci$ as shown in striped region of the inset of \fig{fig_ft}.
This range of quality corresponds to movies with quality above $\avg{Q}_\re$,
where their popularity shows an immediate drop after introduction
and rises afterwards 
(see for instance the lines in \fig{fig_ft} 
with $Q_\alpha/\avg{Q}_\re=1.1$ and  $1.2$).
Physically,
at the first step after introduction
the immediate drop in popularity is induced by the high trendiness,
together with competitions with existing good movies.
At the second step,
the popularity rises as 
a new movie which is usually of lower quality is reviewed by the viewers.
Such behaviors are observed in simulations and real data of Netflix in \fig{gr_netflix},
a website for movie recommendation.

These hits are formed by the successful spreading in the recommender systems,
as similar to other forms of cascade, 
such as information cascade \cite{devany2001}.
Early viewers post their positive reviews and attract more viewers,
reinforcing the ``good gets richer" effect \cite{caldarelli2002}.
After the hits,
many users watched the movie and its on-list popularity ultimately drops.

We expect such hits in box-office,
either at the first or second step,
to be the main cause for the gross box-office boom in \fig{fig_kq}.
Despite a small change in $Q_\alpha/\avg{Q}_\re$ from $0.9$ to $1.1$,
a comparison of $\fa(t)$ in \fig{fig_ft} 
reveals that the hits result in large $\sum_{t}\fa(t)$,
which is consistent with the result shown in \fig{fig_kq}.
We thus expect that $Q_c$ for boom
coincide with $\qcii$ for hits.
These results show that the outcomes of the competitions with on-list movies are crucial in box-office dynamics
and gross box-office.
Defeating existing movies in reviews creates hits and lead to boom,
while losing the competition suppresses popularity and results in small box-office.

\subsection{Competition between good movies}
\label{sec_comp}

Slight modifications of \req{eq_fa} allow us to write down the 
box-office dynamics of two movies introduced one after another,
and enable us to study more directly the competition.
We consider movie $\beta$ introduced 
after movie $\alpha$,
i.e. $t_\beta=t_\alpha+1$.
As the derivation is similar to the that of \req{eq_fa},
the coupled equations for $\fa(t)$ and $\fb(t)$ are given in the appendix.

\begin{figure}
\centerline{\epsfig{figure=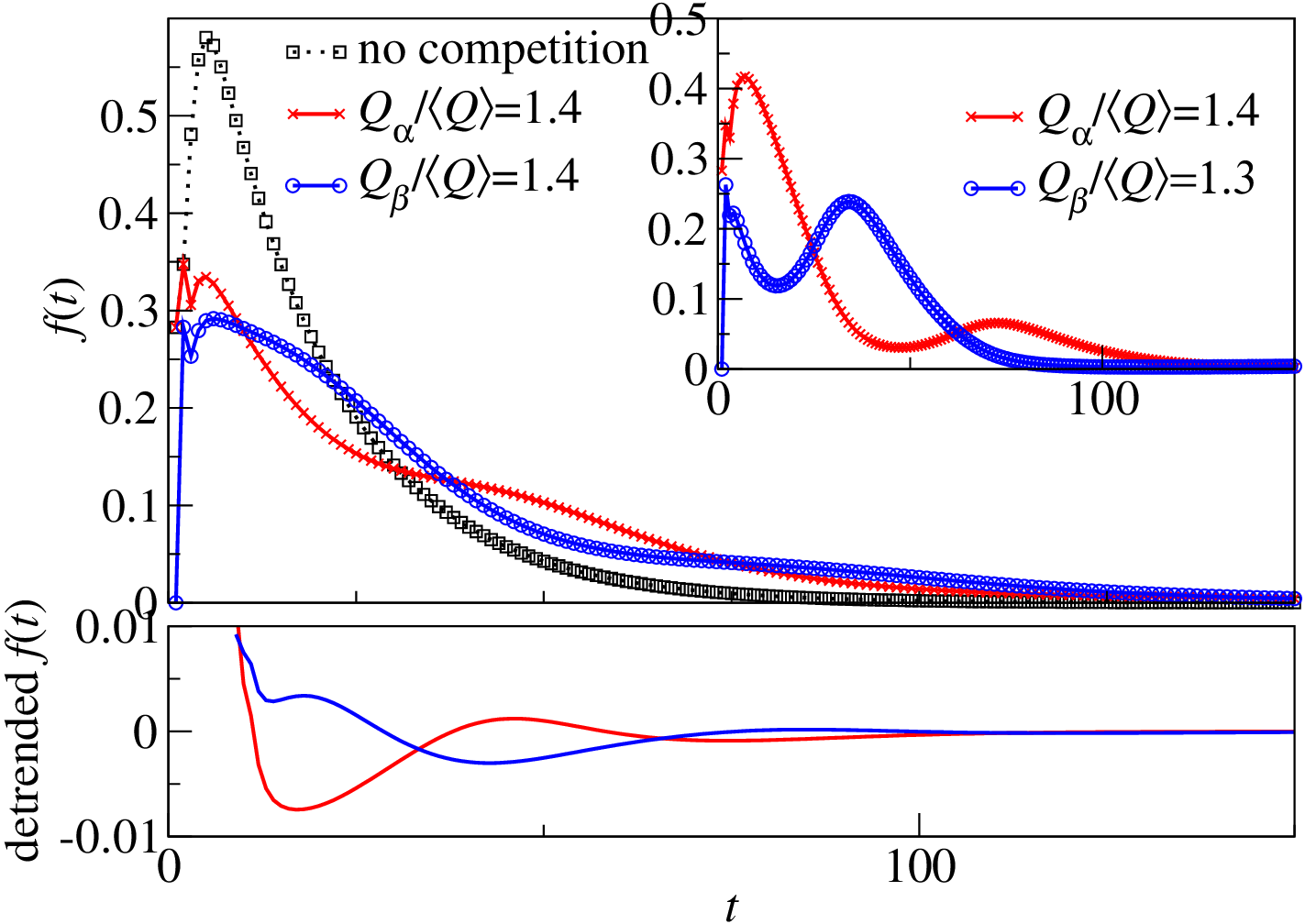, width=0.9\linewidth}}
\caption{
(Color online) Upper penal: $\fa(t)$ and $\fb(t)$ as a function of $t$ for movie $\alpha$ and $\beta$
respectively introduced at time $t_\alpha$ and $t_\alpha+1$ with quality $Q_\alpha=Q_\beta=1.4\avg{Q}_\re$.
Parameters: $p=0.05$, $\tre=0.4$, $\sen=1$ and $\gen=1$.
$\avg{Q}_\re$ are obtained with $\rho(Q)\sim Q^{-2}$ with $\qm=50$.
Inset: The same figure with $Q_\alpha=1.5\avg{Q}_\re$ and $Q_\beta=1.4\avg{Q}_\re$.
Lower penal: detrended $\fa(t)$ and $\fb(t)$,
by subtraction of moving average.
}
\label{fig_comp}
\end{figure}

We first examine the competition between two good movies with 
equal quality.
As shown in the upper penal of \fig{fig_comp},
with $Q_\alpha=Q_\beta=1.4\avg{Q}_\re$,
both $\fa(t)$ and $\fb(t)$ show small oscillations intervening
with each other.
A simple detrending,
e.g. subtraction of moving average (lower penal of \fig{fig_comp}),
reveals that their
box offices are anti-correlated,
implying that they are competing for viewers
through the review list.
As compared to the case with a single movie in the mean-field approximation,
both $\fa(t)$ and $\fb(t)$ show a small or no hit
shortly after its introduction,
but their popularities last longer as they have higher tails.
Remarkably,
we found that the gross box-office of the competing movies 
show only a small drop when compared to its counterpart in the single movie case,
as the tails are long for both $\fa(t)$ and $\fb(t)$.
Such behaviors are not observed in finite size systems,
as small $\fa(t)$ and $\fb(t)$ in the tail is not sufficient to spread popularity,
when $k_\alpha(t), k_\beta(t)<1$.

For movies of identical quality,
we can simplify \req{eq_fa2} to get a simple relation between 
$\fa(t)$ and $\fb(t)$ when $t>t_\alpha+2$,
as given by
\begin{eqnarray}
\label{eq_comp}
	\frac{\fa(t)}{\fb(t)}=\left(\frac{1-p\sum_{t'=0}^{t-1}\fa(t')}{1-p\sum_{t'=0}^{t-1}\fb(t')}\right)\frac{\fa(t-1)}{\fb(t-1)}.
\end{eqnarray}
The relation shows that during the competition,
popularity is influenced by two factors:
(1) the number of accumulated popularity $\sum_{t'=0}^{t-1}f(t')$,
and (2) the popularity $f(t-1)$ in the previous step.
Physically,
it implies that when two movies of equal quality are competing, 
the one which accumulated less box-office get a slight bias in popularity,
as a higher portion of agents did not watch the movie.
This increases the popularity of the biased movie at this step and has a reinforcing effect for its popularity
in the next step.
However,
such effect does not last forever as the number of accumulated popularity ultimately increases
and an opposite trend starts.
This phenomenon causes a balancing effect on the popularity of the competing movies,
and results in intervening between the movies' popularity,
leading to similar gross box-office for the two movies.
Such intervening is more prominent when the two movies differ in their quality,
as shown in the inset of \fig{fig_comp}.

\section{Comparison with Simulation and Empirical Results}
\label{sec_sim}

Finally,
we compare our theoretical mean-field approximation
with simulation and empirical results,
and look at details which are not captured by the mean-field approach.

\begin{figure}
\centerline{\epsfig{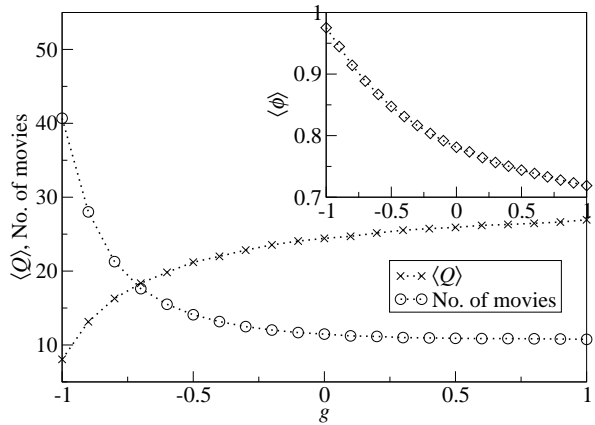}}
\caption{
The average quality and the number of movies on the recommender list in simulations,
as a function of $\gen$.
Parameters: $p=0.05$, $\tre=0.5$, $\sen=1$ and $\gen=1$.
Quality of movies is drawn from $\rho(Q)\sim Q^{-2}$ with $\qm=50$.
Inset: the fraction of relevant reviews as a function of $\gen$.
}
\label{fig_simQ}
\end{figure}

In \fig{fig_simQ},
we show the average quality of movies 
and the number of movies on the review list in simulations.
As expected,
when $\gen$ increases,
the quality of movies on the list increases.
However,
the number of on-list movies decreases:
choice becomes more limited
when reviewers tend to review good movies.
It corresponds to a trade-off of quality with diversity,
which is found in other recommendation systems \cite{tao2010}.
Note that $\avg{Q}$ corresponds to the average over all on-list movies,
which is different from $\avg{Q}_\re$ in the mean-field approximation.
Nevertheless, 
$\avg{Q}_\re$ shows the same trend as $\avg{Q}$.

We further define the fraction of on-list reviews from movies
which have not been watched by a potential viewer $i$ at time $t$ to be $\phi_i(t)$,
given by
\begin{eqnarray}
	\phi_i(t) = 1-\sum_{r=0}^L \theta(t-t_{m_r}+2) a_{i, m_r}.
\end{eqnarray}
The average value of $\phi$ over users and time is shown 
in the inset of \fig{fig_simQ},
As $\gen$ increases,
$\avg{\phi}$ decreases which show that when diversity 
decreases,
viewers are more likely to find a watched movie on the list.
This makes the recommender system less effective.

\begin{figure}
\centerline{\epsfig{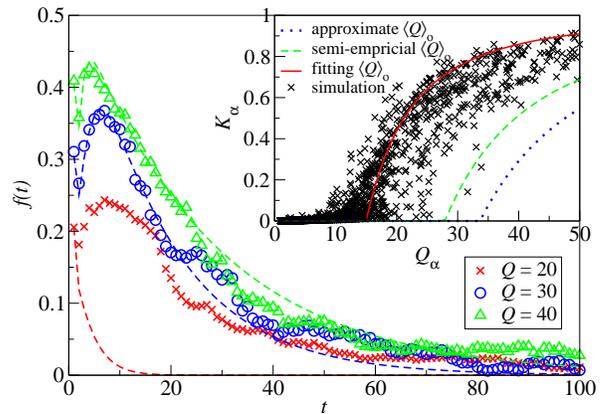}}
\caption{
(Color online)
The popularity $f(t)$ from simulations (symbols) and mean-field approximation 
with semi-empirical $\avg{Q}_\re$ (dashed lines).
Parameters: $p=0.05$, $\tre=0.5$, $\sen=1$ and $\gen=1$.
Quality of movies is drawn from $\rho(Q)\sim Q^{-2}$ with $\qm=50$.
Inset: gross box-office as a function of movie quality.
The corresponding values of $\avg{Q}_\re$ for
analytic,
semi-empirical and fitting
are 33.7, 
27.9
and 15.
}
\label{fig_simkq}
\end{figure}

In the inset of \fig{fig_simkq},
we then compare the $K-Q$ relation between the mean-field results (\ref{eq_grossApprox2}) and 
the simulation results.
Despite the fit with the approximate $\avg{Q}_\re$ (from \req{eq_avgQapprox}) does not agree
with the simulation results,
it does capture the features of the drastic increase in $K_\alpha$
when $Q_\alpha$ is beyond a critical quality.
We thus incorporate the simulated values of $\avg{Q}_\re$
in \req{eq_grossApprox2},
which is shown by the green dashed line as semi-empirical prediction.
Though capturing the trend of the drastic rise,
the prediction is below most of the data points.
We have examined the origin of the discrepancy by comparing $\pi_{i,\alpha}(t)$ in simulations
and the predicted values in the mean-field approximation,
which shows that averaging $\pi_{i,\alpha}(t)$ instead of only its denominator
(as in the present approach) would yield a better approximation.
Nevertheless,
incorporating a smaller $\avg{Q}_\re$ than the simulated value gives the 
red solid line in the inset of \fig{fig_simkq},
which agrees well with simulations.
It implies that the present approach captures the major features of the model,
given an estimate of effective $\avg{Q}_\re$.
As a result,
competition with the effective $\avg{Q}_\re$ is thus 
the major factor in driving box-office dynamics.

We then compare the simulated with the predicted $f(t)$
incorporated with semi-empirical values of $\avg{Q}_\re$ in \fig{fig_simkq}.
We have improved the mean-field prediction by incorporating
the simulated $\avg{\phi}$ as a coefficient for $\avg{Q^{g+1}}_\re$ in \req{eq_meanpi}.
The simulated $f(t)$'s are obtained by averaging box-office
of movies within a small range centered at the indicated $Q_\alpha$.
As can been seen,
the predicted $f(t)$ fits well for large value of $Q$,
and the simulated $f(t)$ show also the hits starting from the second step.
We remind readers that physical origins of such hits 
are discussed in Section \ref{sec_dyn}.
These simulation results also show that 
the hits in dynamics is the reason for gross box-office boom,
as predicted from our analysis.

\begin{figure}
\centerline{\epsfig{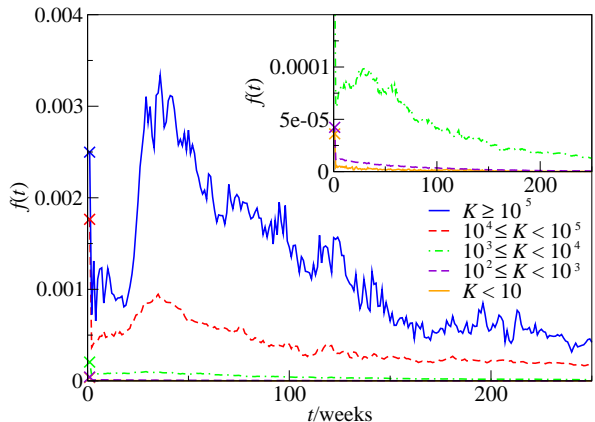}}
\caption{
(Color online) The popularity $f(t)$ from 5000 movies on Netflix in the period from Oct 1998 to Dec 2005,
with $K$,
the total number of viewers as indicated in the range.
The number of active users  is $4.8\times 10^5$ in this period.
The symbol $\times$ indicate the average weekly ``opening" of movies.
}
\label{gr_netflix}
\end{figure}

Finally we show the popularity $f(t)$ of movies as obtained from Netflix.
The number of reviewers $N_r$ on Netflix increases with time due to the its increasing popularity,
we thus put $f_\alpha(t)=k_\alpha(t)/N_r(t)$ which corresponds to the share of reviews on movie $\alpha$ at time $t$.
As the intrinsic quality of movies are not known,
we suggest to distinguish movie quality by the number of total viewers $K$,
and plot in \fig{gr_netflix} the average $f_\alpha(t)$ from movie $\alpha$ with $K_\alpha$ falling in a particular range.
From \fig{gr_netflix},
the two upper curves show a high ``opening" and an immediate drop in the second weeks,
while a prominent hit occurs afterwards.
They correspond to $f(t)$ from movies with large $K$,
i.e. movies of high quality.
For movies with small $K$,
the inset of \fig{gr_netflix} shows in expanded vertical scale that
hit is only observed for movies with $10^3\le K<10^4$,
but not for movies in the groups of lower $K$.
These empirical results show qualitative agreements with the results obtained in the present model,
which suggest the validity of the present description for the fundamental box-office dynamics.

\section{Conclusion}
\label{sec_con}

We studied the competition of movies through reviews in a simple model
of recommender system.
By adopting the mean-field approximation for movie interaction,
we show that,
for movies defeating the average review score,
their popularity spreads through the review systems
as similar to other forms of cascade.
Popularity hits are formed either at the first or second steps
after the introduction of these movies,
and result in booms in gross box-office.
The average review score thus characterizes the critical quality 
of movies to become blockbusters.
Such average score represents the average quality of peer competitors
which implies hits are more likely to occur 
when competitors are bad movies. 
On the other hand,
less generous reviewers and
low intention for watching new movies
create more prominent hits.
These results show that the outcomes of the competitions with on-list movies are crucial in box-office dynamics
and gross box-office.
Defeating existing movies in reviews creates hits and lead to boom,
while losing the competition suppresses popularity and results in small box-office.

Generalizing the mean-field approximation allow us to analyze and show that the 
box offices of two competing good movies are anti-correlated,
which produce long lasting awareness as compared to the case of single good movies.
The model reveals the significant role of movie competition in understanding box-office dynamics.

We remark that the model can be modified to study dynamics 
of movie viewers.
For example,
more reviews can be shown on the list by storing reviews of older than one day,
according to popularity 
movie quality or a suitable decay function.
Agents and movies of heterogeneous taste and attribute can be modelled,
while review scores are given according to the corresponding overlap.
Such modifications may reveal more fundamental aspects and interesting dynamics driving
movie box-office.

\section*{Acknowledgments}

We thank Stanislao Gualdi for very fruitful discussions.
This work is partially supported by 
the Liquid Publications and QLectives projects (EU FET-Open
Grants 213360 and 231200).

\appendix

\section{Dynamical equations of correlated popularity}

In this appendix,
we write down the dynamical equations for $\fa(t)$ and $\fb(t)$
discussed in Section \ref{sec_comp},
regarding the competition between movie $\alpha$ and $\beta$
introduced one after the other.
The coupled equations are given by
\begin{widetext}
\begin{eqnarray}
\label{eq_fa2}
	f_\alpha(t) &=&
	\begin{cases}
	\displaystyle(1-\tre)[1-p\fa(0)]\left[1+\frac{1-\tre}{\fa(0)}\left(\frac{\avg{Q}_\re}{Q_\alpha}\right)^{\gen+1}\right]^{-1}
	&\mbox{for $t=t_\alpha+1$,}
	\\
	\displaystyle(1-\tre)\left[1-p\sum_{t'=0}^{t-1}\fa(t')\right]
	\left[1+\frac{f_\beta(0)}{\fa(t-1)}\left(\frac{Q_\beta}{Q_\alpha}\right)^{\gen+1}
	+\left(\frac{1-\tre}{\fa(t-1)}-1\right)\left(\frac{\avg{Q}_\re}{Q_\alpha}\right)^{\gen+1}\right]^{-1}
	&\mbox{for $t=t_\alpha+2$.}
	\\
	\displaystyle(1-\tre)\left[1-p\sum_{t'=0}^{t-1}\fa(t')\right]
	\left[1+\frac{f_\newm(0)}{\fa(t-1)}\left(\frac{Q_\newm}{Q_\alpha}\right)^{\gen+1}
	+\frac{\fb(t-1)}{\fa(t-1)}\Bigg(\frac{Q_\beta}{Q_\alpha}\right)^{\gen+1}
	\\
	\displaystyle
	\quad\quad\quad
	\quad\quad\quad
	\quad\quad\quad
	\quad\quad\quad
	\quad\quad\quad
	\quad\quad\quad
	+\left(\frac{1-\tre-\fb(t-1)}{\fa(t-1)}-1\right)\left(\frac{\avg{Q}_\re}{Q_\alpha}\right)^{\gen+1}\Bigg]^{-1}
	&\mbox{for $t>t_\alpha+2$.}
	\end{cases}
	\\
	\nonumber\\
	\fb(t) &=& 
	\begin{cases}
	\displaystyle(1-\tre)[1-p\fb(0)]\left[1+\frac{\fa(t-1)}{\fb(0)}\left(\frac{\avg{Q}_\alpha}{Q_\beta}\right)^{\gen+1}
	+\left(\frac{1-\tre-\fa(t-1)}{\fb(0)}\right)\left(\frac{\avg{Q}_\re}{Q_\alpha}\right)^{\gen+1}\right]^{-1}
	&\mbox{for $t=t_\alpha+2$,}
	\\
	\displaystyle(1-\tre)\left[1-p\sum_{t'=0}^{t-1}\fb(t')\right]
	\left[1+\frac{f_\newm(0)}{\fb(t-1)}\left(\frac{Q_\newm}{Q_\beta}\right)^{\gen+1}
	+\frac{\fa(t-1)}{\fb(t-1)}\Bigg(\frac{Q_\alpha}{Q_\beta}\right)^{\gen+1}
	\\
	\displaystyle
	\quad\quad\quad
	\quad\quad\quad
	\quad\quad\quad
	\quad\quad\quad
	\quad\quad\quad
	\quad\quad\quad
	+\left(\frac{1-\tre-\fa(t-1)}{\fb(t-1)}-1\right)\left(\frac{\avg{Q}_\re}{Q_\beta}\right)^{\gen+1}\Bigg]^{-1}
	&\mbox{for $t>t_\alpha+2$.}
	\end{cases}
\end{eqnarray}
\end{widetext}


\end{document}